\documentclass[aip,
 amsmath,amssymb,
 reprint,%
]{revtex4-1}

\usepackage{graphicx}
\usepackage{dcolumn}
\usepackage{bm}

\usepackage[utf8]{inputenc}
\usepackage[T1]{fontenc}
\usepackage{mathptmx}
\usepackage{etoolbox}
\usepackage{parskip}
\usepackage{siunitx}

\makeatletter
\def\@email#1#2{%
 \endgroup
 \patchcmd{\titleblock@produce}
  {\frontmatter@RRAPformat}
  {\frontmatter@RRAPformat{\produce@RRAP{*#1\href{mailto:#2}{#2}}}\frontmatter@RRAPformat}
  {}{}
}%
\makeatother
\begin{document}

\preprint{}

\title[]{Physical Properties and Thermal Stability of Zirconium Platinum Nitride Thin Films}
\author{R.A. Gallivan}
 \affiliation{Laboratory for Nanometallurgy, Department of Materials, ETH Zurich}
 \email{rebecca.gallivan@mat.ethz.ch}
 \author{J. Manser}
 \affiliation{Laboratory for Nanometallurgy, Department of Materials, ETH Zurich}
\author{A. Michelini}
 \affiliation{Laboratory for Nanometallurgy, Department of Materials, ETH Zurich}
 \author{N. Toncich}
 \affiliation{Laboratory for Nanometallurgy, Department of Materials, ETH Zurich}
 \author{N. Abando Beldarrain}
 \affiliation{Laboratory for Nanometallurgy, Department of Materials, ETH Zurich}
\author{C. Vockenhuber}
 \affiliation{Ion Beam Physics, Department of Physics, ETH Zurich}
\author{A. M\"uller}
\affiliation{Ion Beam Physics, Department of Physics, ETH Zurich}%

\author{H. Galinski}
\affiliation{Laboratory for Nanometallurgy, Department of Materials, ETH Zurich}%
 \email{henning.galinski@mat.ethz.ch}
 \homepage{http://met.mat.ethz.ch/}

\date{\today}

\begin{abstract}
Ternary transition metal nitrides (TMNs) promise to significantly expand the material design space by opening new functionality and enhancing existing properties. However, most systems have only been investigated computationally and limited understanding of their stabilizing mechanisms restricts translation to experimental synthesis. To better elucidate key factors in designing ternary TMNs, we experimentally fabricate and analyze the physical properties of the ternary Zr-Pt-N system. Structural analysis and DFT modeling demonstrate that Pt substitutes nitrogen on the non-metallic sublattice, which destabilizes the rock-salt structure and forms a complex cubic phase. We also show insolubility of Pt in the Zr-Pt-N at 45 at\% with the formation of a secondary Pt-rich phase. The measured reduced plasma frequency, decrease in resistivity, and decrease in hardness reflect a dominance of metallic behavior in bonding. Contrary to previous computational predictions, Zr-Pt-N films are shown to be metastable systems where even low Pt concentrations (1\%) facilitate 
a solid reaction with the Si-substrate, that is inaccessible in ZrN films. 
\end{abstract}

\maketitle

Transition metal nitrides (TMNs) are a peculiar class of interstitial compounds, where nitrogen forms a non-metallic sublattice on the interstitial sites of the crystal. These materials can assume metallic, ionic, and covalent behavior at the same time because of their unique combination of a finite density of states (DOS) at the Fermi energy, wave-function overlap, and electrostatic interactions between the metallic and non-metallic sublattice. Depending on the DOS, stoichiometry, and defects in these sublattices, TMNs can reach diamond-like hardness~\cite{MENG2012_ZrNHardness3,Staia2019_ZrNHardness1,TUNG2009_ZrNHardness2}, high melting-points~\cite{JEHN1988_ZrN_Properties,WriedtMurray_TiNProperties}, plasmonic dispersion~\cite{boltasseva2011low}, high carrier mobility~\cite{adachi2005thermal,MuensterResistance}, catalytic activity~\cite{Attfield2020,https://doi.org/10.1002/adfm.202300623}, magneto-ionic cyclability~\cite{10.1063/5.0079762} and superconductivity~\cite{Potjan2023_superconducting}. This versatility coupled with the CMOS-compatibilty of nitrides such as ZrN, TiN, and HfN offers great potential for conventional and quantum applications.
\par
While binary TMNs are widely studied, ternary TMNs remain largely unexplored. Existing experimental work like that on ZnZrN~\cite{Woods-Robinson2022_ZnZrN} and TiAlN~\cite{Schnabel2018_TiAlN} highlight a key challenge of producing stable rather than metastable ternary TMNs. Furthermore the large potential alloying space and limited insights to stabilizing mechanisms complicate materials selection for designing new ternary TMNs. Recent advances in high throughput material screening~\cite{JAIN20112295} help circumvent these challenges and enable density functional theory (DFT) calculated stability maps like the one introduced by Sun \textit{et al.}~\cite{Sun2019_ternaryMetalNitrides}. Although these DFT simulations provide promising evidence for the stability of a large variety of ternary metal nitrides~\cite{Sun2019_ternaryMetalNitrides}, few of these systems have been experimentally fabricated or characterized. 
\par
In particular, the Zr-Pt-N system shows promise for functional enhancements in catalytic properties by the incorporation of a highly active metal like platinum. ZrN itself has strong catalytic properties with a recent study showing that ZrN nanoparticles can outperform Pt in alkaline solutions~\cite{Attfield2020}. Similarly Zr-oxynitrides demonstrate strong catalytic activity as shown through oxygen reduction reaction (ORR) measurements in acid media~\cite{doi:10.1021/acsomega.6b00555,doi:10.1021/acsaem.7b00100} and exhibit high oxygen diffusivity~\cite{kilo2008fast}. These properties make them a good candidate for a mixed ionic-electronic conductor. Thus, introducing platinum, a highly active metal, may provide even greater functional enhancements which are critical to applications from thermophotovoltaics to fuel cells and plasmon enhanced chemistry. Despite its potential as a functional material, Zr$_4$Pt$_2$N's structure and properties have only been reported computationally~\cite{NPt2Zr4} and no experimental synthesized Zr-Pt-N systems are known. 
\par
In this study, we fabricate and explore the ternary nitride Zr-Pt-N with special emphasis on the effect of an increasing Pt-content on phase stability and optical, mechanical, and electrical properties. Through magnetron sputtering, we deposit ZrPt$_x$N$_y$ as films with varying concentrations of Pt. The structure and functional properties of these new Zr-Pt-N films are compared to the binary nitride counterpart, ZrN, a common nitridic material with a high melting temperature ($2900^{\circ}$C)~\cite{JEHN1988_ZrN_Properties}, high mechanical durability and hardness (23 to 30~GPa)~\cite{MENG2012_ZrNHardness3,Staia2019_ZrNHardness1,TUNG2009_ZrNHardness2}, and similar electrical resistivity to its metallic counterpart~\cite{adachi2005thermal,MuensterResistance}.  

\begin{figure*}[t!]
    \centering
    \includegraphics[width=0.95\textwidth]{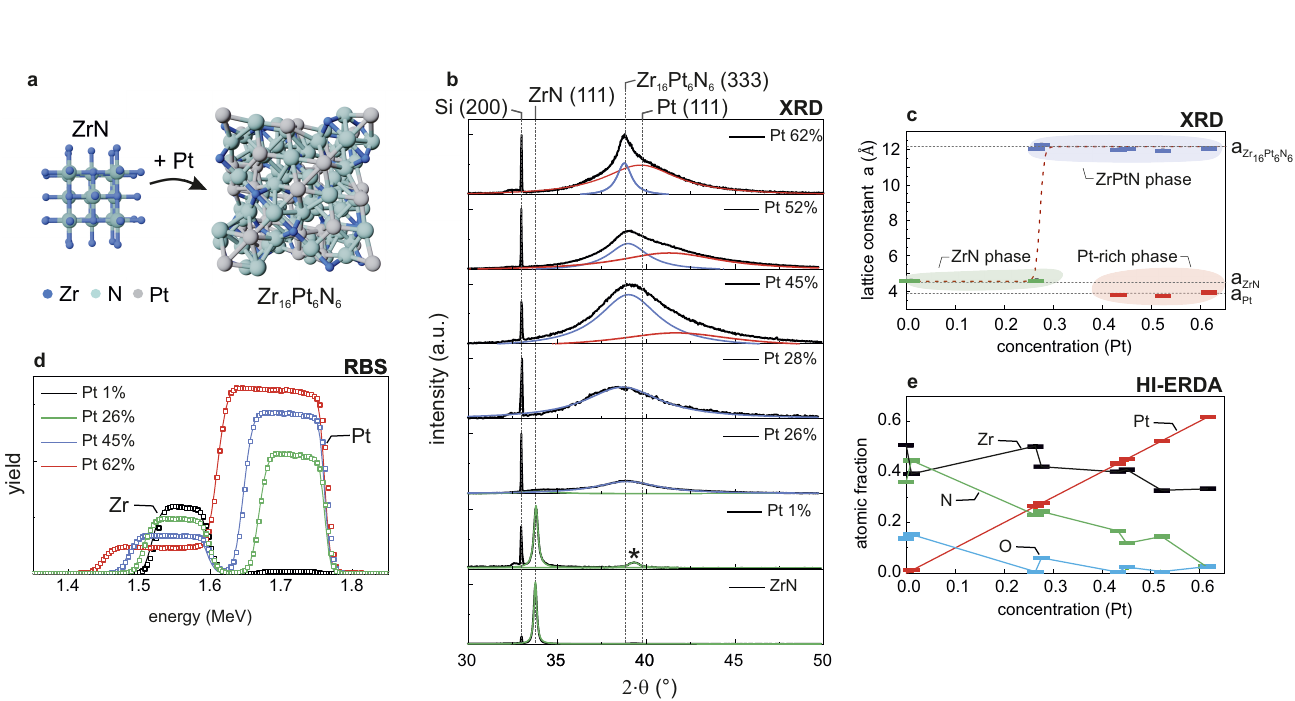}
    \caption{Phase Formation. (a) Transition of the crystal structure from ZrN (rock-salt) to Zr$_{16}$Pt$_6$N$_6$ (diamond-like) due to the addition of Pt. Depicted crystal structures are based on fully-relaxed DFT calculations. (b) XRD spectra of ZrPt$_x$N$_y$ thin films with increasing Pt concentration ($0-62$\%). Deconvoluted peaks correspond to the ZrN-phase (green), the ZrPtN-phase (blue) and secondary Pt nuclei (red). The annotated Zr$_{16}$Pt$_6$N$_6$ (333) peak has been derived from DFT calculations. $*$ denotes the (200) belonging to ZrN. Co-existence of ZrN phase and Zr$_{16}$Pt$_6$N$_6$ occurs at Pt$=26$\%. (c) Calculated lattice constants as a function of Pt concentration. Literature/DFT-based values are annotated at the right side.(d) RBS spectra of selected ZrPtN films with increasing Pt-content showing high uniformity. (e) Evolution of the film chemistry as function of Pt concentration based on elastic recoil detection analysis with heavy ions (HI-ERDA).}
    \label{fig:1}
\end{figure*} 

To link the evolution of the physical properties to changes in chemistry, it is important to get insights into the atomic structure of these complex ternary nitrides. For this purpose, it is convenient to assume mono-crystalline and defect-free materials as a reference since their atomic structure is computationally available using first-principle techniques. Figure~\ref{fig:1}a depicts the optimized crystal structure of ZrN and Zr$_{16}$Pt$_6$N$_6$ calculated with Quantum Espresso~\cite{Giannozzi2009QUANTUMMaterials}. Here, ZrN forms a rock-salt structure with four atoms per unit cell. With the addition of Pt atoms, the unit cell expands to $28$ atoms/cell and forms a stable diamond-like structure. This aligns with previous DFT calculations reported in the literature~\cite{NPt2Zr4,Sun2019_ternaryMetalNitrides}.
\par
The experimental structural and chemical data based on Rutherford backscattering (RBS)~\cite{Schatz1992}, X-ray diffraction (XRD), and elastic recoil detection analysis with heavy ions (HI-ERDA)~\cite{Schatz1992} presented in Figure~\ref{fig:1} identify the formation of a distinct ZrPt$_x$N$_y$ phase where Pt substitutes on the N sublattice sites. At low Pt concentrations
($\leq 1$\%) the film maintains the rock-salt structure typical of
ZrN (green peak) and simply incorporates Pt as a defect (Figure~\ref{fig:1}b).
However, at larger concentrations of Pt, the ZrPt$_x$N$_y$ peak (blue) appears and the material shifts to
a new cubic unit cell with a lattice parameter of $12$~\AA (Figure~\ref{fig:1}c). For films with a high concentration of Pt ($\geq$45$\%$), a second peak emerges (red). Based on the lattice parameter calculations highlighted in Figure~\ref{fig:1}c, this broad peak is associated with a distinct Pt-rich phase. The extensive peak broadening suggests that the Pt forms very small ($<1$~nm) and finely dispersed crystalline nuclei.  
\par
As Pt is introduced, it substitutes on the N site due to the strong
bonding interaction between Zr and Pt~\cite{WangANDCarter1993_PtZrBond}. While predicted computationally, this Pt substitution of N is also observed experimentally using HI-ERDA as increases in Pt content correspond to equivalent decreases in N content (Figure~\ref{fig:1}e). Zr remains at a fairly
consistent atomic fraction independent of Pt concentration. Oxygen content also decreases from 0.15 to less than 0.03 atomic fraction with increasing Pt content. The oxygen content observed is mostly attributed to surface oxidization based on the localization observed in the HI-ERDA results (see Supplementary Information).
RBS measurements shown in Figure~\ref{fig:1}d confirm the chemical uniformity of the sputter deposited films.

\begin{figure*}[t!]
    \centering
      \includegraphics[width=0.95\textwidth]{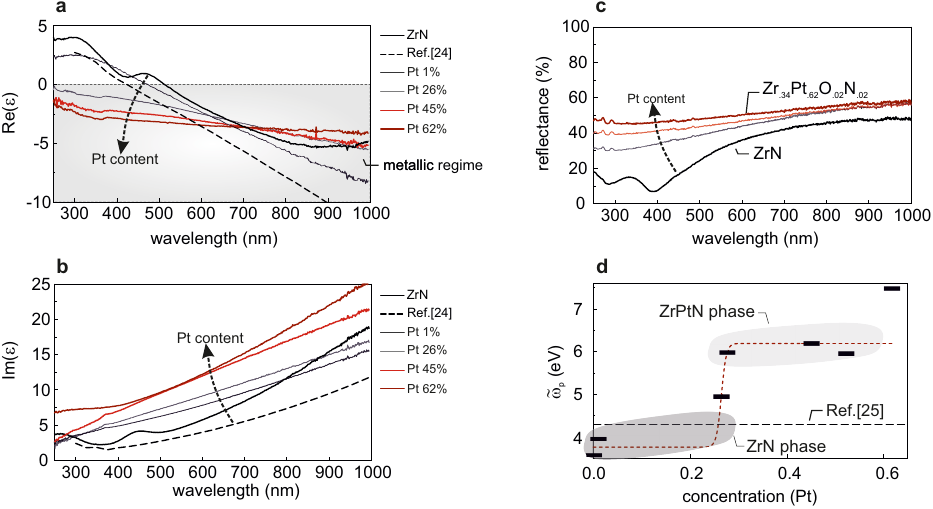}
    \caption{Optical Properties. (a) Real and (b) imaginary part of the dielectric function for five examined ZrPt$_x$N$_y$ films in the as-deposited state and compared to ZrN data from literature~\cite{ribbing_-_1997}. Substitution of N with Pt on the nitrogen sublattice continuously blue-shifts the metallic regime i.e.   Re($\epsilon$)<0. (c) Near-normal incidence reflectance spectra of four selected ZrPt$_x$N$_y$ films with increasing Pt-content. (d) Reduced plasma frequency ($\tilde{\omega}_p$) versus Pt-content. $\tilde{\omega}_p$ is obtained by fitting the experimentally measured permittivity from (a) and (b) using a Drude-Lorentz model. The step-like increase reflects both, the significant increase in free carrier concentration with increasing Pt-content and the crystallographic phase-transition from observed by XRD. The effective plasma frequency of ZrN from Ref.~\cite{https://doi.org/10.1002/adma.201205076} is plotted for comparison.}
    \label{fig:2}
\end{figure*}

We characterize the optical response of the ternary metal nitrides for different Pt concentration via ellipsometry and near-normal incidence reflectometry. Figure~\ref{fig:2} highlights the control in optical property through the alloying of Pt in these ternary nitridic films. Addition of Pt flattens the real part of the film's permittivity $\epsilon$ and shifts the crossover frequency to the metallic regime, i.e. $\Re{\epsilon(\omega_p)}$, further into the UV (Figure~\ref{fig:2}a). Simultaneously, the imaginary part of the permittivity, i.e. the optical loss, increases as function of Pt content for all wavelengths (Figure~\ref{fig:2}b). The extension of the metallic regime is also featured in the materials reflectance (Figure~\ref{fig:2}c). Under near-normal reflectance, the addition of Pt increases reflectance at all wavelengths but with particular increase at wavelengths below 600~nm. For Pt concentrations $\geq$26\%, the films form a broadband reflector. While the response of ZrN is still dominated by interband transition that lead to the formation of a characteristic absorption edge and goldish coloration~\cite{Kumar:16}, the addition of Pt changes the band structure and consequently the optical transitions significantly. 
\par
To quantify this transition, we parameterize the complex permittivity with a finite set of Drude-Lorentz (DL) oscillators~\cite{10.1063/1.1979470}. The DL model takes into account the effective mass $m^{*}$ of the charge carriers and is suited to represent interband transitions. The crossover frequency to the metallic regime is given by the reduced plasma $\tilde{\omega}_p=\omega{_p}/\sqrt{\epsilon_\infty}=\sqrt{n e^{2}/m^{*}\epsilon_\infty}$, which mainly depends on the ratio of the background permittivity $\epsilon_\infty$ and the density of carriers $n$ taking part in the transition. Figure~\ref{fig:1}d presents the screened plasma frequency $\tilde{\omega}_p$ as a function of Pt concentration. For ZrPt$_x$N$_y$ phases, $\tilde{\omega}_p$ increases more than 2-fold from 3.6~eV in ZrN to 5.0~eV in 26\% Pt, 6.2~eV in 45\% Pt, and 7.5~eV in 62\% Pt. We interpret the observed shift to higher energies as result of an overall increase in carrier concentration $n$, which is also reflected in the calculated total density of states (see Supplementary Material). The step-like transition in the range 0-52\% Pt of the reduced plasma frequencies $\tilde{\omega}_p$ can be attributed to the observed crystallographic phase transition (Figure~\ref{fig:1}) that drastically alters the band-structure of the material. For 62\% Pt, the measured $\tilde{\omega}_p$ increases significantly, pointing to an increased contribution of the secondary Pt-phase to the total physical properties. Using the general relation between the optical loss and the plasma frequency $(2/\pi)\int_{0}^{\infty}\omega \text{Im}({\epsilon})\mathrm{d\omega}= \omega_{p}^{2}$~\cite{bassani1976electronic}, one can observe that also the measured optical loss (Figure~\ref{fig:2}b) features this phase transition, suggesting an overall increase in metallic bonding in the material. 

\begin{figure*}[t!]
    \centering
      \includegraphics[width=0.95\textwidth]{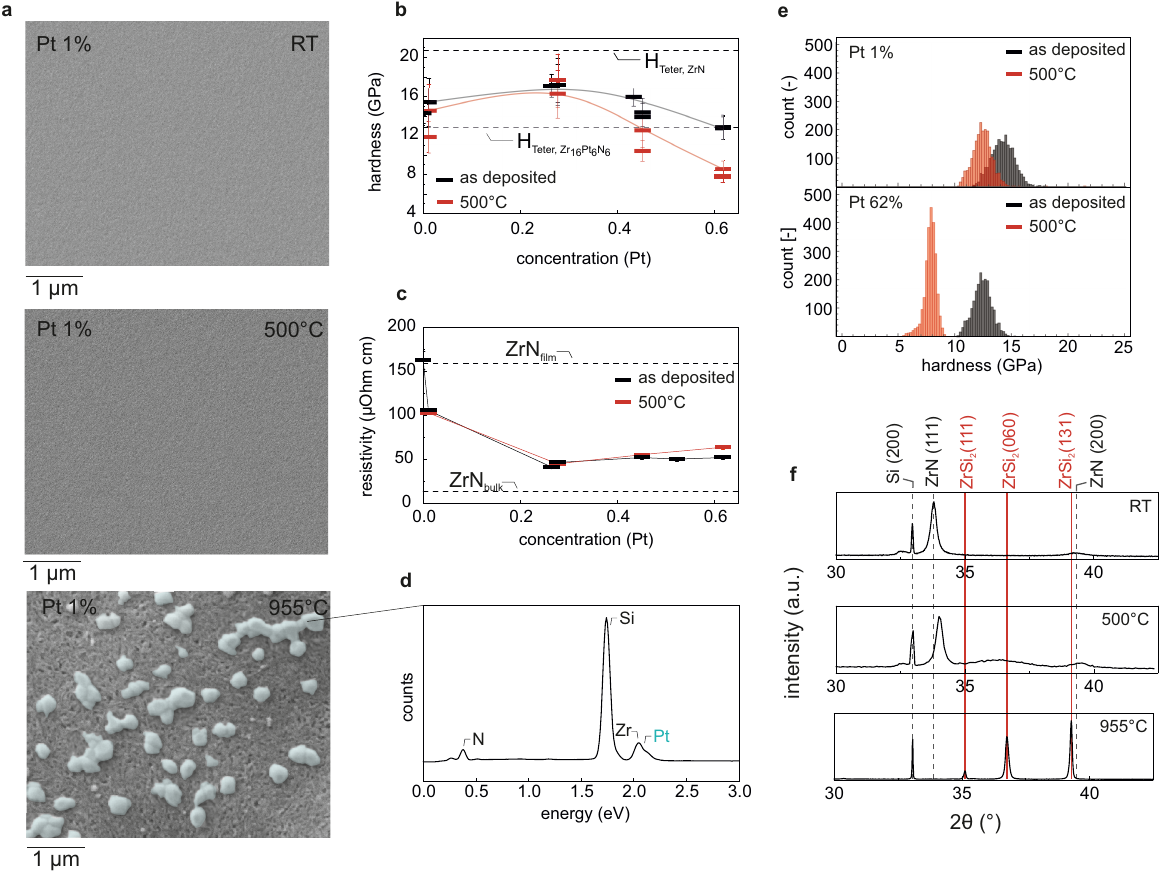}
    \caption{Hardness, Resistivity, and Thermal Stability. (a) SEM images of a 1\% Pt film at room temperature, 500~$^{\circ}$C, and 955~$^{\circ}$C. Pt precipitates are false-colored in blue. (b) Nanoindentation hardness of films as a function of Pt for room temperature (black) and 500 $^{\circ}$C (red). Comparison made to reference Teter hardness values for ZrN and Zr$_16$Pt$_6$N$_6$ estimated from DFT calculations. (c) Resistivity of films as a function of Pt concentration for room temperature (black) and 500 $^{\circ}$C (red) as compared to both film \cite{Signore_2010} and bulk \cite{Wang1995} resistivity of ZrN. (d) EDS spectra highlighting the presence of Pt in SEM image of the 955 $^{\circ}$C annealed sample. (e) Histograms of nanoindentation data for 1\% Pt film (top) and 62\% Pt film (bottom) both as deposited (black) and after annealing at 500~$^{\circ}$C (red). (f) XRD spectra for 1\% Pt at room temperature, 500~$^{\circ}$C, and 955~$^{\circ}$C.}

    \label{fig:3}
\end{figure*}

Having established the structural and optical properties of the Zr-Pt-N system, we now address the thermal stability of this new ternary TMN phase. Figure~\ref{fig:3}a shows a set of scanning electron micrographs obtained from the Zr-Pt-N with smallest Pt concentration, namely $1$\%. Up to 500~$^{\circ}$C, the ZrPt$_x$N$_y$ films show no morphological change. As-deposited films also exhibit comparable hardness to other ZrNs, particularly zirconium oxy-nitrides~\cite{Staia2019_ZrNHardness1,TUNG2009_ZrNHardness2,MENG2012_ZrNHardness3} (Figure~\ref{fig:3}b), and a reduced resistivity compared to ZrN films~\cite{Signore_2010} (Figure~\ref{fig:3}c). 
\par
Histograms of the indentation data Figure~\ref{fig:3}e show generally Gaussian distributions of hardness with no apparent clustering or bimodal presentation both before and after annealing. The hardness peaks at 26~at\% Pt (Figure~\ref{fig:3}b), which can be explained both by a shift from strengthening through Pt replacement of oxygen to soften from both the subsequent replacement of N with Pt and the growth of a Pt-rich phase. Figure~\ref{fig:1}e highlights the reduction of oxygen content to negligible levels with the incorporation of Pt. As Zr-O bonds require the occupation of antibonding states electronically, these bonds are inherently less stable than either Zr-N or Zr-Pt bonds and typically promote vacancy formation to help stabilize the lattice structure~\cite{Schwandt1997}. Pt$_3$Zr structures form highly stable bonds~\cite{WangANDCarter1993_PtZrBond} which drive preferential incorporation of Pt over Zr-O bonds and mechanistically describes the strengthening effect observed. However, incorporation of Pt onto the N sites (detailed in Supplementary) reduce the hardness as compared to pure ZrN and the emergence of a nanoscale Pt-rich phase observed (Figure~\ref{fig:1}b) act as softening mechanism. Since Pt's hardness (H$_v$ = 9.4~GPa~\cite{FarrorANDMclellan1977_PtShear}) is lower than ZrPt$_x$N$_y$, the nanoscale Pt-rich nuclei lower the film's average hardness and create localized opportunities for deformation. 
\par
While resistivity remains unaffected (Figure~\ref{fig:3}c) at 500~$^{\circ}$C, clear structural degradation has begun and by 955~$^{\circ}$C, chemical and structural decomposition is visible(Figure~\ref{fig:3}a, d, f). Precipitate formation on the film's surface (false-colored blue) is confirmed to be Pt particles by EDS (Figure~\ref{fig:3}d). Furthermore, XRD measurements emphasize the structural degradation into a mixture of Pt, Si from the substrate, and ZrSi$_2$ phases which are only present above 500~$^{\circ}$C (Figure~\ref{fig:3}f). The mechanical degradation can also be explained by formation and growth of the Pt-rich phase. While surface precipitation of large Pt-rich particles is observed when annealed at 900~$^{\circ}$C (Figure~\ref{fig:3}a), these phases likely nucleate within the film at lower temperatures. This destabilization of the ZrPt$_x$N$_y$ films does not impact the dominant metallic bonding character and thus does not influence the observed electrical behavior.
 \par
This evidence for phase segregation and low thermal stability in ZrPt$_x$N$_y$ ternary alloys stand in contrast to DFT calculated stability maps~\cite{Sun2019_ternaryMetalNitrides}. Critically, we observe a solid state reaction between Pt-containing ZrN and the Si substrate when annealed at 900~$^{\circ}$C (Figure~\ref{fig:3}f). This reaction completely consumes the ZrN phase and forms a crystalline ZrSi$_2$. While pure films of Zr and Si can react at 500-600~$^{\circ}$~\cite{Yamaychi1991_ZrSi1,Tanaka1995_ZrSi2}, ZrN-Si interfaces are stable up to 900~$^{\circ}$C~\cite{Ruan2009_ZrNStability, Takeyama2000_SivsZrNStability}. As shown by XRD in both Figure~\ref{fig:1}b and \ref{fig:3}f, 1\% Pt films are structurally ZrN with incorporated Pt defects in the rocksalt lattice structure. Thus the complete conversion of the film to a ZrSi$_2$ phase at 900~$^{\circ}$C underscores Pt's key role in destabilizing the ZrN structure and catalyzing the solid-state Zr-Si reaction. This decomposition of the ZrPt$_x$N$_y$ film emphasizes its metastability.  
\par
Together these experiments emphasize the dominance of Zr-Pt metallic bonding in Zr-Pt-N alloys. It drives the formation of the metastable ZrPt$_x$N$_y$ films through substitution of Pt on the N sublattice and transformation to a large, open unit cell structure. When extending to the design of other ternary TMNs, it is important to consider the metal-metal interactions in comparison to the parent metal nitride bonds. While DFT sets a key basis for establishing promising directions for ternary TMN design, this work emphasizes the critical role of experimental validation and investigation to fully identify destabilizing factors and address feasibility of ternary metal nitrides for functional applications. 
%

\end{document}